\documentclass[iop,apjl]{emulateapj}
\usepackage{amsmath}
\usepackage{rotating}

\shorttitle{VERY-LIGHT INTERNALLY-SUBSONIC AGN JETS}
\shortauthors{GUO}

\begin{document}
\bibliographystyle{apj} 

\title {On the Importance of Very-light Internally-subsonic AGN Jets in Radio-mode AGN Feedback}

\author{Fulai Guo}

\affil{Key Laboratory for Research in Galaxies and Cosmology, Shanghai Astronomical Observatory, Chinese Academy of Sciences, 80 Nandan Road, Shanghai 200030, China; fulai@shao.ac.cn}

\begin{abstract}

Radio-mode AGN feedback plays a key role in the evolution of galaxy groups and clusters. Its physical origin lies in the kpc-scale interaction of AGN jets with the intracluster medium. Large-scale jet simulations often initiate \emph{light} internally-supersonic jets with density contrast $0.01<\eta<1$. Here we argue for the first time for the importance of \emph{very-light} ($\eta<0.01$) internally-subsonic jets. We investigated the shapes of young X-ray cavities produced in a suite of hydrodynamic simulations, and found that bottom-wide cavities are always produced by internally-subsonic jets, while internally-supersonic jets inflate cylindrical, center-wide, or top-wide cavities. We found examples of real cavities with shapes analogous to those inflated in our simulations by internally-subsonic and internally-supersonic jets, suggesting a dichotomy of AGN jets according to their internal Mach numbers. We further studied the long-term cavity evolution, and found that old cavities resulted from \emph{light} jets spread along the jet direction, while those produced by \emph{very light} jets are significantly elongated along the perpendicular direction. The northwestern ghost cavity in Perseus is pancake-shaped, providing tentative evidence for the existence of very light jets. Our simulations show that very-light internally-subsonic jets decelerate faster and rise much slower in the ICM than light internally-supersonic jets, possibly depositing a larger fraction of jet energy to cluster cores and alleviating the problem of low coupling efficiencies found previously. The internal Mach number points to the jet's energy content, and internally-subsonic jets are energetically dominated by non-kinetic energy, such as thermal energy, cosmic rays, or magnetic fields.

\end{abstract}


\section{Introduction}
\label{section:intro}

Radio-mode AGN feedback is widely thought to play a key role in the evolution of hot gas and massive elliptical galaxies in groups and clusters of galaxies, suppressing cooling flows and the associated growth of massive galaxies (e.g., \citealt{mcnamara07}; \citealt{mcnamara12}; \citealt{yuan14}). One compelling evidence comes from numerous detections of kpc-sized surface brightness depressions in X-ray images of galaxy groups and clusters, the so-called ``X-ray cavities". Many cavities are associated with radio jets and spatially coincident with radio lobes (e.g., \citealt{boehringer93}; \citealt{fabian02}; \citealt{birzan04}; \citealt{croston11}), indicating that they are evolved from the interaction of AGN jets with the intracluster medium (ICM). 

AGN jets are accelerated near the vicinities of accreting supermassive black holes (SMBHs), extracting energy from either SMBHs' spin energy \citep{bz77} or accretion disks \citep{bp82}. While they are often observed to have relativistic speeds on pc and smaller scales \citep{shklovskii64}, AGN jets in galaxy groups and clusters often decelerate significantly to form kpc-sized X-ray cavities on kpc to tens-of-kpc scales \citep{mcnamara07}. Some old ghost cavities even seem to rise buoyantly in the ICM, without any signatures of sustaining influence of initial jet momentum (e.g., in the Perseus cluster; \citealt{fabian00}). 

The whole process of jet acceleration, propagation, and deceleration from black hole vicinities (the Schwarzschild radius $\sim 10^{-4}$ pc $M_{\rm bh}/10^{9}M_{\sun}$) to tens of kpc scales is far from clear, involving a radial dynamic range too large to be studied self-consistently in numerical simulations. A related issue is the dominant composition within AGN jets and X-ray cavities, which are quite unclear either (e.g., \citealt{croston14}). State-of-the-art simulations of AGN jets in galaxy clusters typically use spatial resolutions of hundreds pc to few kpc, and often adopt kinetic-energy-dominated jets (the so-called ``mechanical feedback"; e.g., \citealt{gaspari11} and \citealt{yang16}). These jets are supersonic with respect to both the ambient ICM and the internal jet plasma, depositing a large amount of their energy to large distances with relatively low efficiencies in heating cool cluster cores (the ``dentist drill" effect; \citealt{vernaleo06}). 

Recent simulations often invoke jet precession with an angle of around $10^{\circ}$-$25^{\circ}$ to remedy this problem (\citealt{gaspari12}; \citealt{li15}; \citealt{yang16}; \citealt{nawaz16}). Jets with even larger precession angles ($30^{\circ}$-$70^{\circ}$; \citealt{sternberg08a}), wide jets (outflows) with large opening angles (\citealt{sternberg07}; \citealt{gilkis12}; \citealt{prasad15}; \citealt{hillel16}), cosmic-ray-dominated jets \citep{guo11}, and internally transonic jets (\citealt{mendygral12}; \citealt{li14}) have also been studied. Recent observations by \citet{randall15} suggest that shock waves induced by AGN outbursts can heat the ICM roughly isotropically, while theoretical studies have proposed mixing (\citealt{gilkis12}; \citealt{hillel16}), turbulent heating \citep{zhura14}, sound waves \citep{ruszkowski04} as the main channel of transferring the jet energy to the ICM. 

Without resolving the jet propagation on sub-pc scales, kpc-scale properties of AGN jets adopted in numerical simulations may play a key role in predicting the coupling between the jet energy and cool cluster cores. In \citet{guo15}, we used a suite of hydrodynamic simulations to study the connection between jet properties and the shapes of young X-ray cavities, identifying two key jet parameters affecting the cavity shape: density contrast ($\eta\equiv \rho_{\rm jet}/\rho_{\rm amb}$, where $\rho_{\rm jet}$ and $\rho_{\rm amb}$ are the densities of the jet plasma and ambient ICM at the jet base $\sim 1$ kpc) and the internal Mach number ($M_{\rm int} \equiv v_{\rm jet}/c_{\rm s,jet}$, where $v_{\rm jet}$ and $c_{\rm s,jet}$ are the jet speed and the sound speed in the jet interior). The former classifies AGN jets into two types -- internally subsonic ($M_{\rm int} <1$) and internally supersonic ($M_{\rm int} >1$) jets, while the latter may separate jets into three different types:  \emph{heavy} ($\eta>1$), \emph{light} ($0.01<\eta<1$), and \emph{very light} ($\eta<0.01$) jets.

In this paper, we further argue for the physical importance of very-light internally-subsonic AGN jets in radio-mode AGN feedback, with the main purpose of drawing attention of computationalists who almost always adopt light internally-supersonic jets in simulations of AGN feedback in galaxy clusters. In Section 2, we summarize the shapes of young X-ray cavities in our idealized jet simulations, and by comparing with observations, we provide evidence for a potential dichotomy of internally-subsonic and internally-supersonic jets in real galaxy clusters. We then further study the long-term evolution of four representative types of AGN jets in the ICM in Section 3, arguing that very light jets indeed exist in real clusters. In Section 4, we discuss the potential importance of very-light internally-subsonic AGN jets in radio-mode AGN feedback. We summarize and discuss our results in Section 5.

\section{Evidence for Internally Subsonic Jets from Young X-ray Cavities}
\label{section2}

 \begin{figure}
   \centering
\plotone{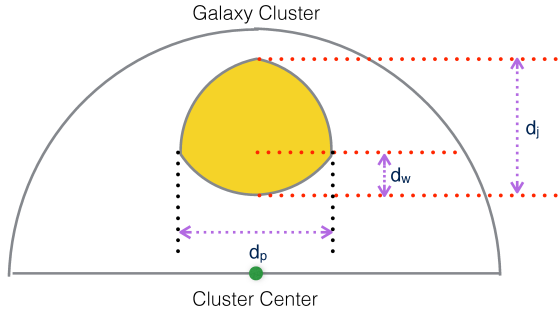} 
\caption{Sketch of an X-ray cavity (yellow-shaded region) in a galaxy cluster. We characterize the cavity shape with two parameters: radial elongation $\tau\equiv d_{\rm j}/d_{\rm p}$ and top wideness $b\equiv d_{\rm w}/d_{\rm j}$. The value of $\tau$ depends sensitively on the projection effect, while $b$ does not.}
 \label{plot1}
 \end{figure}

 \begin{figure}
   \centering
\plotone{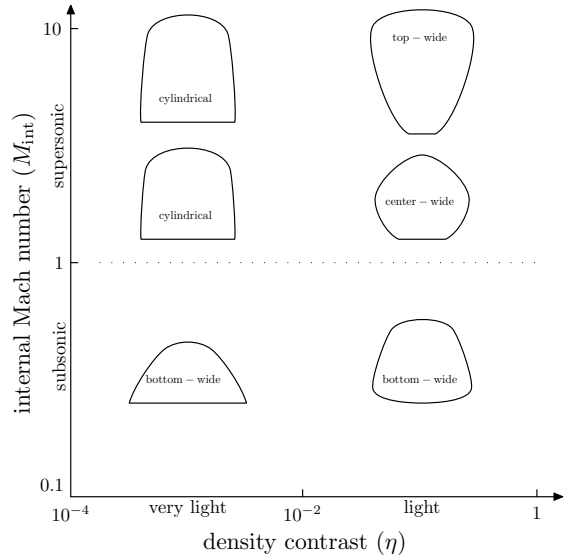} 
\caption{Sketch of young X-ray cavities produced by AGN jets with various values of density contrast $\eta$ and internal Mach number $M_{\rm int}$. The shapes of young X-ray cavities are summarized from hydrodynamical jet simulations in \citet{guo15} and an addition simulation R5 (to produce the right-bottom cavity). The cluster center is assumed to be an arbitrary point below each cavity along its symmetry axis.}
 \label{plot2}
 \end{figure}

The properties of AGN jets strongly affect their propagation and the associated formation of X-ray cavities in galaxy clusters. Using a suite of hydrodynamic jet simulations, \citet{guo15} demonstrates that the shapes of X-ray cavities during and shortly after their formation (thereafter referred as ``young X-ray cavities") are strongly affected by two key jet parameters on kpc scales -- density contrast $\eta$ and the internal Mach number $M_{\rm int}$, while the shapes of old X-ray cavities are further affected by the ICM viscosity. In this section, we further argue that the shapes of young X-ray cavities can be used to probe $M_{\rm int}$, and suggest that at least some AGN jets may be internally subsonic. 

\subsection{Probing Jet Properties with the Shapes of Young X-ray Cavities}

 \begin{figure*}
   \centering
\plottwo{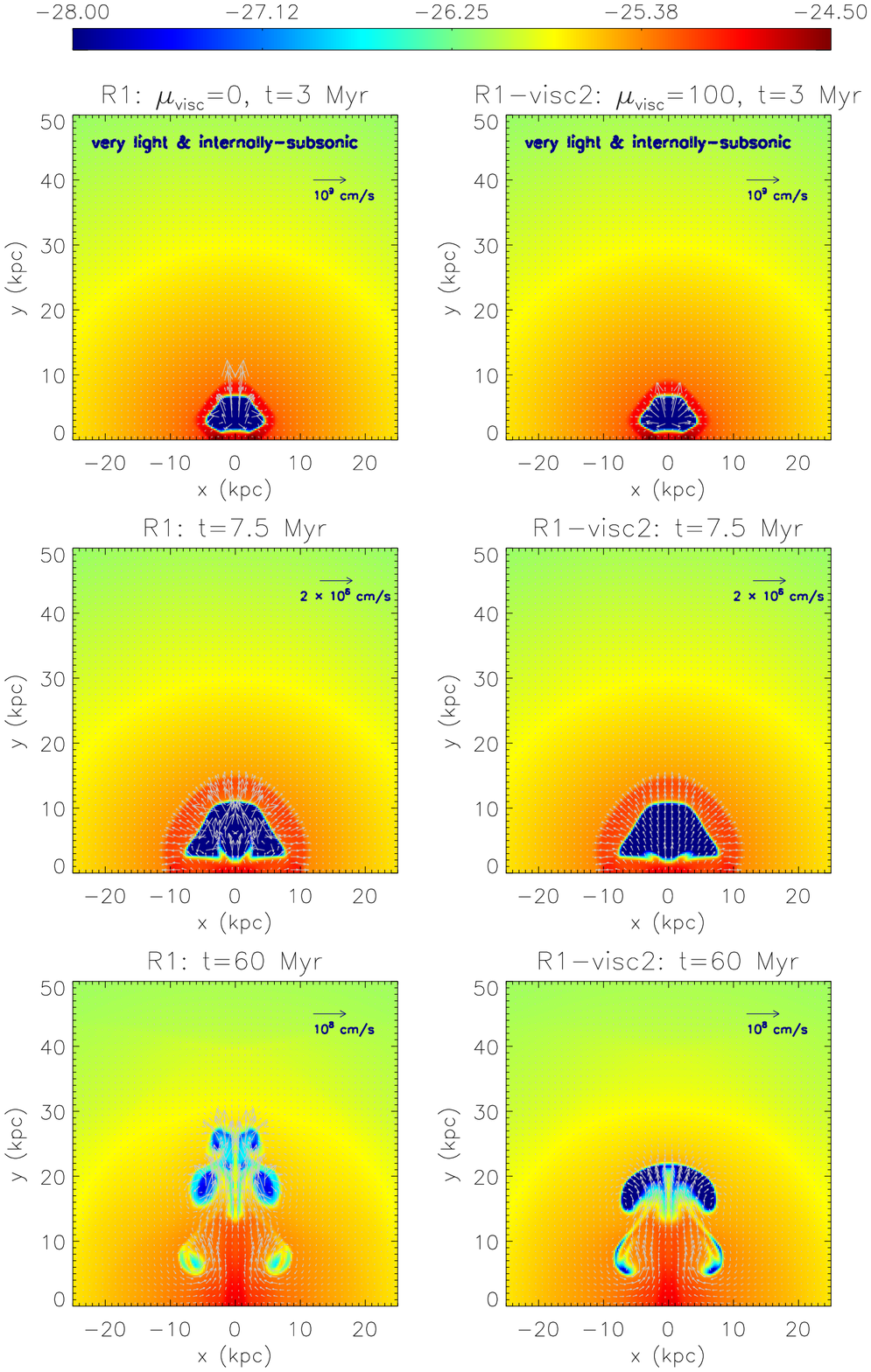} {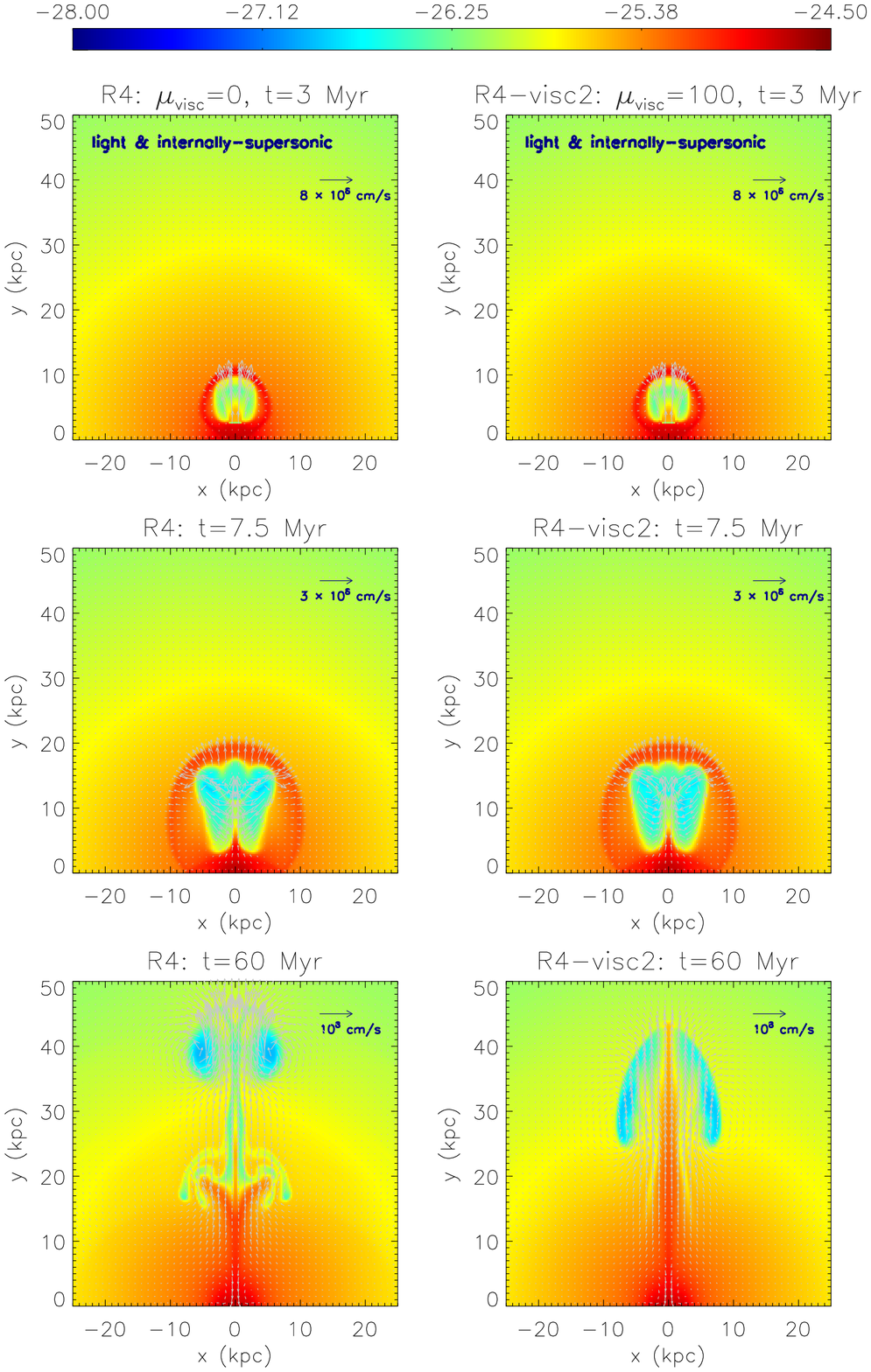}
\caption{Temporal jet evolution in four representative simulations, shown from left to right: a very-light, internally-subsonic jet in the non-viscous ICM (run R1) and viscous ICM (run R1-visc2; $\mu_{\rm visc}=100$ g cm$^{-1}$ s$^{-1}$), and a light internally-supersonic jet in the non-viscous ICM (run R4) and viscous ICM (run R4-visc2). Each panel shows gas density in the logarithmic scale, with arrows denoting gas velocity in the linear scale.}
 \label{plot3}
 \end{figure*} 
 
We investigated jet evolution in a suite of two-dimensional axisymmetric simulations in cylindrical coordinates. The setup of our simulations was presented in detail in \citet{guo15}. We initiate a uniform, well-collimated jet at a distance of around kpc from the cluster center, and follow its evolution in the well-observed Virgo cluster. The jet is active constantly for a period of $t=t_{\rm jet}$, and we focus on the shapes of resulting young X-ray cavities at $t=1.5t_{\rm jet}$ and their long-term evolution in the ICM. Key jet parameters and the ICM dynamic viscosity coefficient ($\mu_{\rm visc}$) in the simulations presented in the paper are listed in Table 1. The jet duration $t_{\rm jet}=5$ Myr and radius $r_{\rm jet}=1$ kpc are the same in all these simulations.

The shapes of X-ray cavities may be characterized by two geometrical parameters, as shown in Figure \ref{plot1}. Radial elongation $\tau\equiv d_{\rm j}/d_{\rm p}$ is defined as the ratio of the cavity axis along the jet direction ($d_{\rm j}$) to that along the direction perpendicular to the jet direction ($d_{\rm p}$). The other parameter, top wideness $b$, represents the relative location in the jet axis of the cavity's widest size along the perpendicular direction, $b\equiv d_{\rm w}/d_{\rm j}$, where $d_{\rm w}$ is the distance from the cavity bottom to the location of its widest  size in the perpendicular direction. From the observational point of view, $b$ is more useful to probe jet properties, as it does not depend on the inclination of the cavity with respect to the line of sight, while the value of $\tau$ is strongly affected by the projection effect. According to the value of $b$, an X-ray cavity can be top-wide ($b>0.5$), center-wide ($b\sim0.5$), or bottom-wide ($b<0.5$).

The dependence of the shapes of young X-ray cavities on $M_{\rm int}$ and $\eta$ is summarized in Figure \ref{plot2}, based mostly on a suite of simulations presented in \citet[see Fig. 2, Fig. 4, and Table 1]{guo15}. The left-top and left-middle panels are based on runs Rv2 and Rv1 in \citet{guo15} respectively, while the left-bottom panel is drawn from results of runs R1 and R2. The right-top and right-middle panels are based on runs R4 and R3 in \citet{guo15} respectively. The right-bottom panel corresponds to a new simulation (denoted as run R5) with $\eta=0.1$ and $M_{\rm int}=0.48$. The setup of run R5 is the same as in run R4, except that the jet energy density is increased by a factor of $\eta_{\rm e}=100$, resulting in a larger sound speed and a smaller Mach number within the jet. The resulting X-ray cavity can be seen in the top-left panel of Figure \ref{plot4}, which shows the synthetic X-ray surface brightness map in run R5 at $t=1.5t_{\rm jet}$.

\begin{table*}
 \centering
 \begin{minipage}{150mm}
  \renewcommand{\thefootnote}{\thempfootnote} 
  \caption{Key Parameters of AGN Jets and the ICM viscosity in the Simulations }
    \vspace{0.1in}
  \begin{tabular}{@{}lccccccccc}
  \hline    
  & {$\eta$\footnote{$\eta=\rho_{\rm jet}/\rho_{\rm amb}$ is the initial jet density normalized by the ambient ICM gas density at the jet base.} }&$M_{\rm int}$& {$\eta_{\rm e}$\footnote{$\eta_{\rm e}=e_{\rm th}/e_{\rm amb}$ is the initial jet thermal energy density normalized by the ambient ICM energy density at the jet base.} } &
         $v_{\rm jet}$& $\mu_{\rm visc}$&$\dot{M}_{\rm jet}$\footnote{$\dot{M}_{\rm jet}=\rho_{\rm jet}\pi r_{\rm jet}^{2}v_{\rm jet}$ is the mass deposition rate that each jet injects into the ICM at the jet base.}&{$P_{\rm ki}$\footnote{$P_{\rm ki}=\rho_{j} v_{\rm jet}^{3}\pi r_{\rm jet}^{2}/2$ is the kinetic power of each jet.}}& {$P_{\rm th}$\footnote{$P_{\rm th}=e_{ \rm j} v_{\rm jet}\pi r_{\rm jet}^{2}$ is the thermal power of each jet.}}& {$P_{\rm tot}$\footnote{$P_{\rm tot}=P_{\rm ki}+P_{\rm th}$ is the total power of each jet.}} \\ Run&
         & &&($10^{9}$cm/s)&($\text{  g cm}^{-1}\text{ s}^{-1}$)&M$_{\sun}$/yr &($10^{44}$erg/s)&($10^{44}$erg/s)&($10^{44}$erg/s) \\ \hline 
         R1 &$0.001$&0.15&$10$& 1 &0&0.0673 &0.02&1.65&1.67 \\ 
              R3 &$0.1$&1.5&$10$& 1 &0& 6.73&2.14&1.65&3.79 \\    
          R4& $0.1$&4.8& $1$ & 1&0& 6.73&2.14  &0.165  &2.30 \\                                  
         R5& $0.1$&0.48& $100$ & 1&0&6.73 &2.14&16.53&18.67 \\
         R1-visc2 &$0.001$&0.15&$10$& 1 &100&0.0673 &0.02&1.65&1.67 \\ 
         R4-visc2& $0.1$&4.8& $1$ & 1&100& 6.73&2.14  &0.165  &2.30 \\       
         Rv1-visc2 &$0.001$&2.2&$1$& 4.64 &100&0.31 &2.13&0.77&2.90 \\ 
         R5-visc2& $0.1$&0.48& $100$ & 1&100&6.73 &2.14&16.53&18.67 \\
          \hline
\label{table1}
\end{tabular}
\end{minipage}
\end{table*}

Figure \ref{plot2} clearly shows the potential of using the shapes of young X-ray cavities, particularly the top wideness, to probe the internal Mach number of AGN jets. Bottom-wide cavities are always produced by internally-subsonic jets, while internally-supersonic jets tend to produce non-bottom-wide cavities, including cylindrical cavities (by very light jets), center-wide and top-wide cavities (by light jets). This result is very robust, and is not affected by the projection effect. As shown in \citet{guo15}, additional jet parameters, e.g., the jet radius and duration, only affect radial elongation of young X-ray cavities, but not top wideness. Combining with deep X-ray observations of galaxy groups and clusters, we encourage X-ray observers to use Figure \ref{plot2} as a guidance to investigate the properties of AGN jets and radio-mode AGN feedback.

Figure \ref{plot3} shows jet evolution in four representative simulations at three representative times $t=3$, $7.5$, and $60$ Myr. The left two columns show the evolution of a very-light internally-subsonic jet in the non-viscous and viscous ICM, while the right two columns show the evolution of a light internally-supersonic jet in the non-viscous and viscous ICM. As expected, the jets in non-viscous runs R1 and R4 produce backflows, and thus form vortices within the cavities. In comparison, the very light jet in run R1 produces stronger backflows, leading to significant side expansion at the bottom of the resulting cavity, and thus resulting in a bottom-wide cavity. Viscosity tends to dissipate velocity shears, and the impact of viscosity on the flow structure is more significant for very light jets ($\partial {\bf v}/\partial t\propto \mu_{\rm visc}/\rho$). As discussed in more detail in \citet{guo15}, for both very light and light jets, viscosity at the level of $\mu_{\rm visc}=100$ g cm$^{-1}$ s$^{-1}$ does not affect the shapes of young X-ray cavities, but significantly affect the long-term cavity evolution by suppressing the development of torus-like structures seen in non-viscous runs.

\subsection{Internally Subsonic Jets and A Dichotomy in Radio-mode AGN Feedback}

 \begin{figure*}
   \centering
\plotone{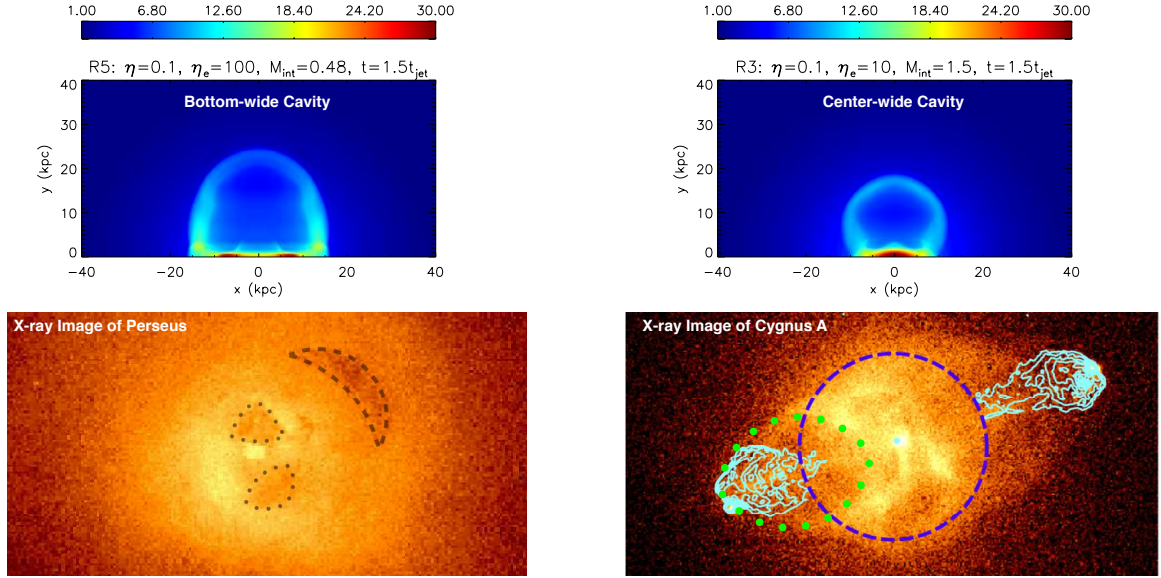} 
\caption{Left panels: A typical bottom-wide X-ray cavity produced by an internally-subsonic jet in the simulation R5 (top) and two inner bottom-wide X-ray cavities (enclosed in dotted lines) in the {\it Chandra} image of Perseus in the 0.5-7 keV energy band (bottom; adapted from Fig. 1 of \citealt{fabian00}). Right panels: A center-wide X-ray cavity produced by an internally-supersonic jet in the simulation R3 (top) and a center-wide X-ray cavity (enclosed in the green dotted line) in the Chandra image of Cygnus A in the 0.5-8 keV energy band (bottom; adapted from Fig. 2 of \citealt{reynolds15}). The contours show the \citet{perley84} 6 cm radio map, which helps determine the East edge of the cavity, while the West edge is determined from the X-ray map directly. The other cavity in the opposite direction is likely also center-wide, though its edge is less clear. The top panels show synthetic X-ray surface brightness maps in our simualtions. The dashed line in the bottom-left panel encloses an old X-ray cavity, while that in the bottom-right panel encloses a circle of 30-kpc radius.}
 \label{plot4}
 \end{figure*} 
 
As indicated in Figure \ref{plot2}, internally-subsonic and internally-supersonic jets produce young X-ray cavities with different shapes in the hot ICM. Two representative examples of young X-ray cavities produced in our simulations are shown in top panels of Figure \ref{plot4}, which shows synthetic X-ray surface brightness maps (line of sight projections of the gas cooling rate along an arbitrary direction perpendicular to the jet axis in units of $10^{-4}$ erg cm$^{-2}$) at $t=1.5t_{\rm jet}=7.5$ Myr. The top-left panel contains a bottom-wide cavity produced by a light, internally-subsonic jet in run R5. Bottom-wide young cavities can also be produced by very light, internally-subsonic jets, as shown in Figure \ref{plot3} (runs R1 and R1-visc2). The top-right panel contains a center-wide cavity produced by a light, internally-supersonic jet in run R3.

The dichotomy of internally-subsonic and internally-supersonic jets in our simulations may also exist in real galaxy clusters. Here we discuss two archetypical AGN feedback events in Perseus and Cygnus A. The bottom-left panel of Figure \ref{plot4} shows the Chandra 0.5-7 keV image of Perseus \citep{fabian00}, which clearly indicates two bottom-wide X-ray cavities enclosed in dotted lines near the center. These two cavities are relatively young, radiating 1.4 GHz radio emissions \citep{fabian00}. According to our scenario (Fig. 2), these two bottom-wide cavities were potentially produced by internally-subsonic jets, which could be light or very light. It was previously argued by \citet{sternberg07} that these two cavities could be produced by wide internally-supersonic jets with a half-opening angle larger than $50^{\circ}$.

On the other hand, the bottom-right panel of Figure \ref{plot4} shows the Chandra X-ray image of Cygnus A \citep{reynolds15}, superposed with two radio lobes (6 cm emission in contours; \citealt{perley84}). The green dotted line encloses a center-wide cavity, whose East (left) edge is constrained from the corresponding radio lobe, while whose West (right) edge is determined from the contrast in X-ray surface brightness. The other cavity in the opposite direction seems also to be center-wide, though its edge is less clear. According to our numerical results (Fig. 2), these center-wide cavities were produced by light, internally-supersonic jets, which can still be seen now in radio \citep{perley84} and X-rays \citep{wilson06}. 

The jets in Cygnus A were previously modeled in \citet{carvalho05} as very-light ($\eta\sim10^{-4}$) internally-supersonic jets. However, \citet{carvalho05} adopted a constant-density atmosphere, while in realistic cluster atmospheres, our simulations \citep{guo15} demonstrated that very-light internally-supersonic jets produce cylindrical cavities, as also seen in \citet{krause05}. The radio lobes in Cygnus A were also studied with cosmic-ray hydrodynamic simulations in \citet{mathews10} and \citet{mathews12}, which however did not directly model jet evolution (but instead using two ``phantom hotspots" to inject cosmic rays).

Thus, Figure 4 suggests that there are two different types of AGN jets in real galaxy clusters: internally subsonic and internally supersonic jets. The internal Mach number $M_{\rm int}$ points to the energy content of AGN jets. The value of $M_{\rm int}$ is connected with the ratio of the jet's kinetic energy density ($e_{\rm kin}=\rho_{\rm jet} v_{\rm jet}^{2}/2$) to thermal energy density ($e_{\rm th}$):
\begin{eqnarray}
\frac{e_{\rm kin}}{e_{\rm th}}=\frac{\gamma(\gamma-1)}{2}M_{\rm int}^{2}{\rm ~,} \label{eq1}
\end{eqnarray}
which is derived from the definition of $M_{\rm int}$:
\begin{eqnarray}
 M_{\rm int}\equiv \frac{v_{\rm jet}}{c_{\rm s,jet}}=\frac{v_{\rm jet}}{\sqrt{\gamma P_{\rm jet}/\rho_{\rm jet}}}{\rm ~.} \nonumber 
\end{eqnarray}
\noindent
Here $P_{\rm jet}=(\gamma-1)e_{\rm th}$ denotes the pressure within the jet, and $\gamma$ is the adiabatic index of the jet material ($\gamma=5/3$ for ideal thermal gas). Thus internally supersonic jets roughly correspond to jets energetically dominated by the kinetic energy, while internally subsonic jets are jets energetically dominated by non-kinetic energy, such as thermal energy, cosmic rays (\citealt{guo08a}; \citealt{guo11}), or magnetic fields \citep{xu08}.

A dichotomy of radio-mode AGN feedback has also been previously seen in radio observations, i.e., Fanaroff-Riley (FR) type I and II radio sources \citep{fr74}. It would be very interesting to investigate the connection between the jet dichotomy suggested in our current study and the FR dichotomy inferred from radio observations, and the origins of both dichotomies. While FR I radio sources seem to prevail over FR II sources in galaxy clusters, an observational study on the relative frequency of internally subsonic and supersonic jets may shed new insights onto radio-mode AGN feedback.

\section{Possible Evidence for Very light Jets from Old X-ray Cavities}
\label{section:old}

 \begin{figure*}
   \centering
\plottwo{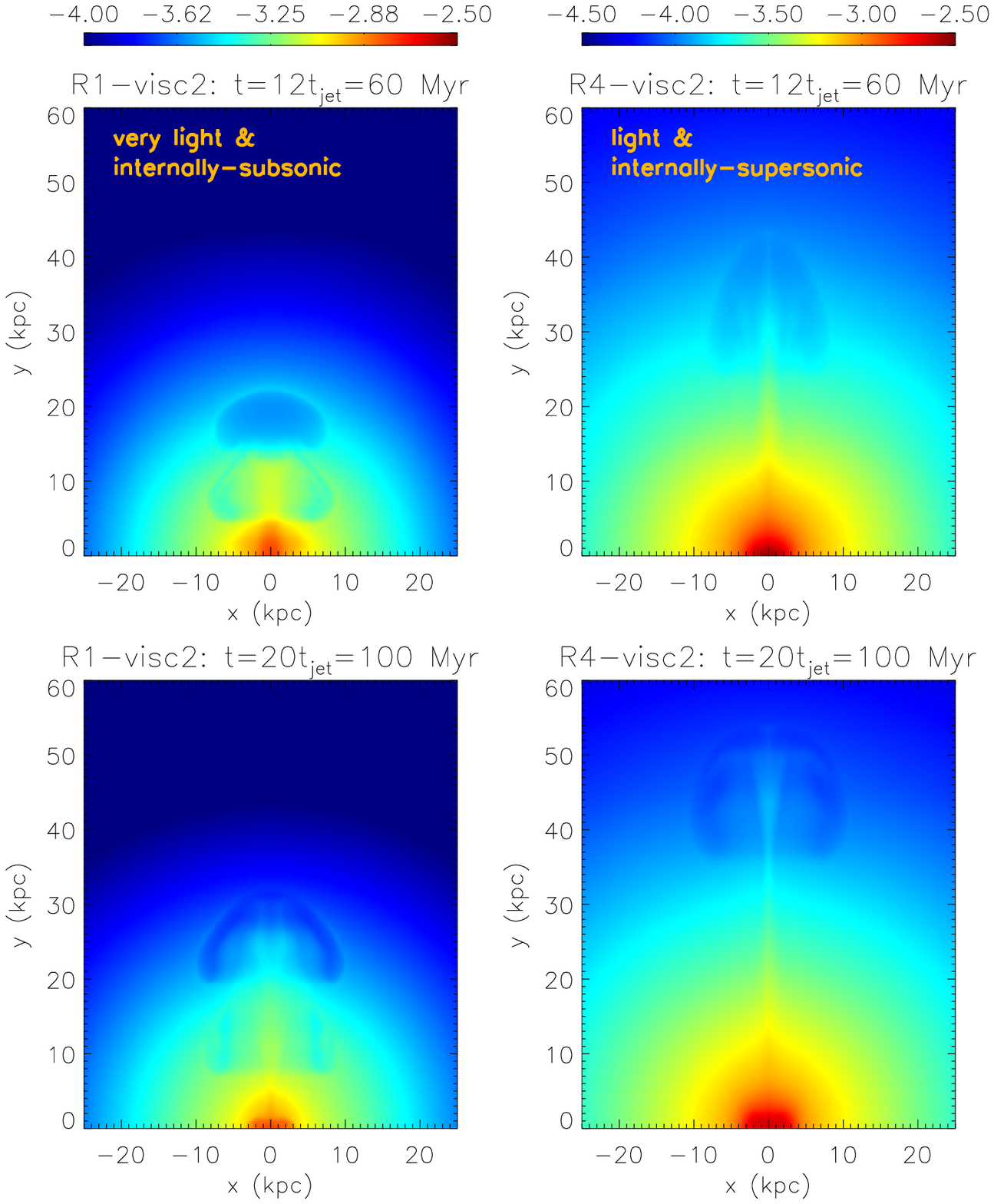} {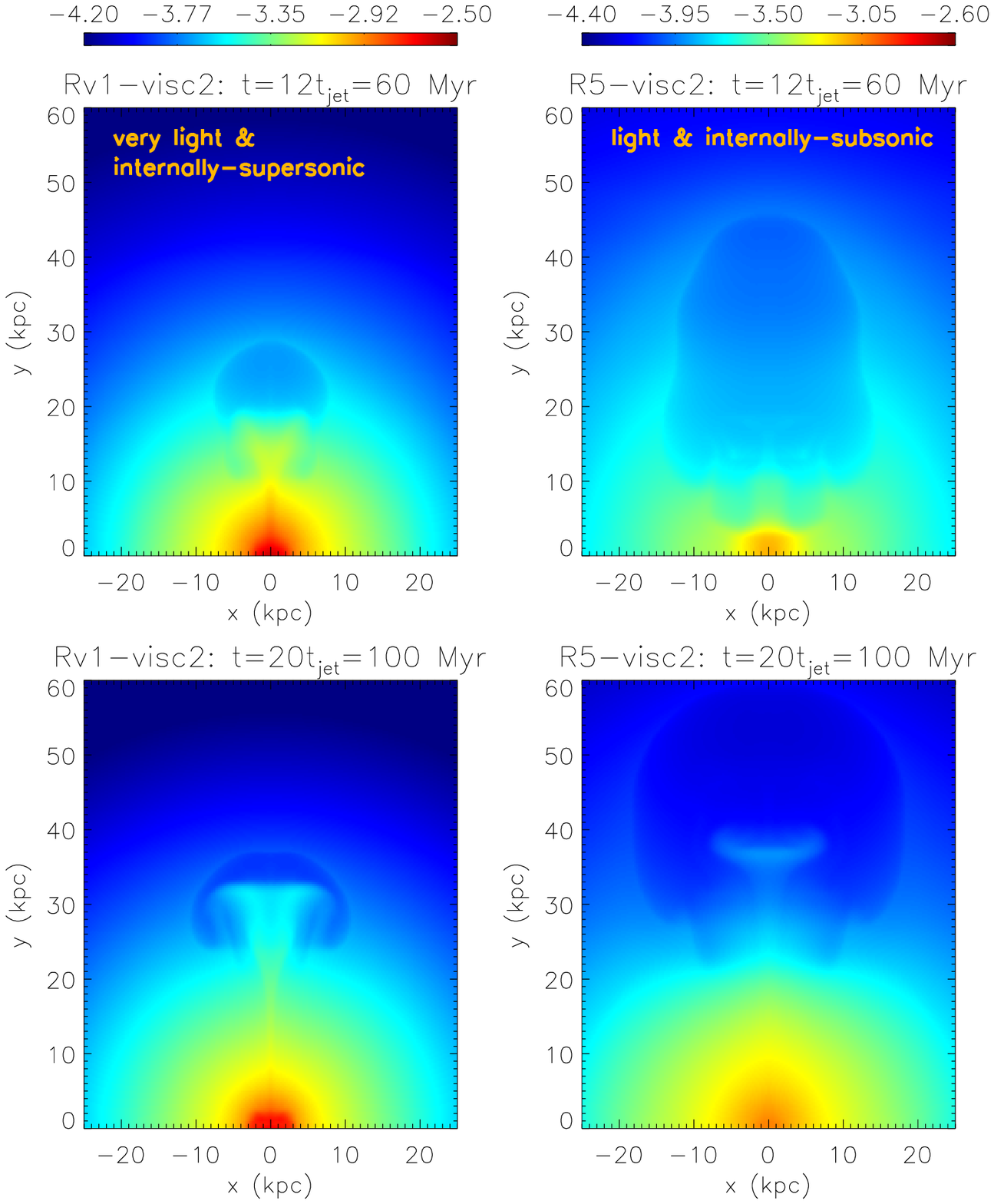}
\caption{Long term evolution of four different types of AGN jets in hydrodynamic simulations, shown from left to right: a very-light, internally-subsonic jet (run R1-visc2), a light internally-supersonic jet (run R4-visc2), a very-light, internally-supersonic jet (run Rv1-visc2), and a light, internally-subsonic jet (run R5-visc2). With color bars chosen to maximize the cavity contrast, the images show synthetic X-ray surface brightness maps in the logarithmic scale at $t=60$ Myr (top panels) and $100$ Myr (bottom panels), while the jet is only active for the first $t_{\rm jet}=5$ Myr.}
 \label{plot5}
 \end{figure*} 

Studies of young X-ray cavities in the previous section suggest that there are two different types of AGN jets according to their internal Mach number. How do they evolve differently in the ICM and what do we learn by comparing their evolution in numerical simulations with observations? This is a more difficult question to probe, as additional physics, such as shear viscosity (\citealt{kaiser05}; \citealt{reynolds05}; \citealt{guo12b}; \citealt{guo15}) and magnetic tension (\citealt{kaiser05}; \citealt{jones05}; \citealt{ruszkowski07}), may play a significant role in the long-term jet evolution. Pure hydrodynamic simulations of initially-static cavities predict that cavities are quickly disrupted by Rayleigh-Taylor (RT) and Kelvin-Helmholtz (KH) instabilities \citep{reynolds05}, while more realistic simulations of cavities directly inflated by AGN jets show that they are more stable to these interface instabilities (\citealt{pizzolato06}; \citealt{sternberg08}), possibly due to strong vortices formed in cavities as seen in Figure \ref{plot3}. In addition, Figure \ref{plot3} further shows that associated with the vortex formation, the major body of the cavity tends to evolve into a torus-like structure, while if a significant level of shear viscosity is present, the torus-like feature is significantly suppressed. 

Assuming that viscosity is not strongly suppressed in the ICM and adopting a constant dynamic viscosity coefficient $\mu_{\rm visc}=100$ g cm$^{-1}$ s$^{-1}$, here we investigate the long-term jet evolution in four representative simulations: R1-visc2 (a very-light, internally-subsonic jet), R4-visc2 (a light, internally-supersonic jet), Rv1-visc2 (a very-light, internally-supersonic jet), and R5-visc2 (a light, internally-subsonic jet). The jet parameters in these runs are listed in Table 1, and the results are presented in Figure 5. With this level of viscosity (a significant fraction of the Braginskii viscosity $\sim156(\text{ln}\Lambda/37)^{-1}(T/2\text{ keV})^{2.5}$ g cm$^{-1}$ s$^{-1}$), RT and KH instabilities are effectively suppressed and cavities do not evolve into prominent torus-like structures seen in non-viscous simulations (e.g., \citealt{reynolds05} and \citealt{guo15}). 

As shown in Figure 5, the shapes of old X-ray cavities produced by these four types of AGN jets show some subtle differences. Old cavities in runs R1-visc2 and Rv1-visc2 appear to be flattened and elongated along the direction perpendicular to the jet direction, while cavities in runs R4-visc2 and R5-visc2 are more spread along the jet direction. This is probably because the jets in the former two runs are very light ($\eta=0.001$), losing momentum in the ICM more quickly than light jets ($\eta=0.1$) in the latter two runs. Interestingly, the outer northwestern X-ray cavity in Perseus (ghost cavity enclosed in the dashed line in the left-bottom panel of Fig. 4) is also significantly flattened (pancake-shaped), bearing morphological similarity to old X-ray cavities produced by very light jets in runs R1-visc2 and Rv1-visc2 at $t=100$ Myr. While the pancake-shaped ghost cavity in Perseus was previously argued to be attributed to RT instabilities \citep{soker02}, in our scenario it instead provides tentative evidence for the existence of very light AGN jets in galaxy clusters.

Both light and very light AGN jets may exist in real clusters, as we argued above that the outer cavity in Perseus was produced by a very light jet, while the jets in Cygnus A are light (see Sec. 3). Previous simulations of radio-mode AGN feedback usually adopted light AGN jets (e.g., \citealt{gaspari11} and \citealt{yang16}), while very few studies explored very light jets (e.g., \citealt{krause03}; \citealt{krause05}; \citealt{guo11}). It is possible that very light jets play an important role in radio mode AGN feedback, as further discussed in the following section.

\section{Potential Importance of Very-light Internally-subsonic Jets in AGN Feedback}
\label{section:4}

In the previous two sections, we provide evidence for the existence of both internally subsonic jets and very light jets in galaxy clusters, based on the shapes of young and old X-ray cavities, respectively. Which further suggests that very-light internally-subsonic jets may exist in real clusters. While light internally-supersonic jets, which may indeed exist in some clusters (as possibly in Cygnus A), have often been adopted in numerical investigations of radio-mode AGN feedback, very-light internally-subsonic jets may play a more important role.

Among various types of AGN jets classified based on density contrast and the internal Mach number, very-light internally-subsonic jets contain the lowest momentum density for a given total energy density $e_{\rm tot}=e_{\rm kin}+e_{\rm th}$. The momentum density of a jet $p_{\rm jet}=\rho_{\rm jet} v_{\rm jet}=(2\rho_{\rm jet} e_{\rm kin})^{1/2}$ may be written as 
\begin{eqnarray}
 p_{\rm jet}=(2\rho_{\rm jet} e_{\rm tot})^{1/2} \left[1-\frac{2}{2+\gamma(\gamma-1)M_{\rm int}^{2}}\right]^{1/2} {\rm ~,}
\end{eqnarray}
where we have used Equation (\ref{eq1}). For a fixed value of $e_{\rm tot}$, a jet with smaller values of $\rho_{\rm jet}=\eta \rho_{\rm atm}$ and $M_{\rm int}$ has a lower momentum density $p_{\rm jet}$, and thus very-light internally-subsonic jets have lower momentum density compared to very-light internally-supersonic, light internally-subsonic, or light internally-supersonic jets.

Due to lower density and momentum, very-light internally-subsonic jets are expected to decelerate faster in the ICM. As seen in Figure 5, the cavity produced by a very-light internally-subsonic jet in run R1-visc2 rises slowest, and its average distance to the cluster center at $t=100$ Myr is around $25$ kpc, corresponding to an average rising speed of $245$ km/s. This speed is substantially smaller than the sound speed in the Virgo's ICM $c_{\rm s}=726(k_{\rm B}T/2 {\rm ~keV})^{1/2}$ km/s, which is often adopted as the cavity rising speed in observational studies (e.g., \citealt{rafferty06}). It is also slightly smaller than the buoyancy rising speed $0.5v_{\rm s}$ estimated by \citet{churazov01}, possibly due to the additional effect of viscosity in our simulation (as also seen in Fig. 7 in \citealt{guo15}).

In particular, very-light internally-subsonic jets rise much slower in the ICM than light internally-supersonic jets often adopted in numerical simulations of radio-mode AGN feedback (e.g., \citealt{gaspari11} and \citealt{yang16}). They spend longer time in inner regions of galaxy clusters, possibly having a higher efficiency of coupling their energy to cool cluster cores (through shock heating, turbulent heating, or mixing) and alleviating the problem of low coupling efficiencies associated with light internally-supersonic jets seen in previous studies \citep{vernaleo06}. A detailed study of this speculation is out of the scope of the present paper.

\section{Summary and Discussion}
\label{section:discussion}

The evolution of hot gas and massive galaxies in galaxy groups and clusters depends strongly on the energy input from central SMBHs via radio-mode AGN feedback, whose effectiveness is potentially affected by the properties of AGN jets on kpc scales. Previous studies have been usually focused on light internally-supersonic jets (e.g., \citealt{gaspari11} and \citealt{yang16}). In this paper, we provide evidence for the existence of both internally subsonic jets and very light jets in real galaxy clusters, by comparing observations of the shapes of young and old X-ray cavities with a series of jet simulations. We further argue for the first time that there may be a dichotomy of radio-mode AGN feedback in galaxy clusters originated from internally-subsonic and internally-supersonic AGN jets, and discuss the potential importance of very-light internally-subsonic jets in AGN feedback.

The shapes of young kpc-sized X-ray cavities resulted from jet interaction with the ICM may be used to probe jet properties. From a suite of ideal jet simulations, we conclude that top-wideness of X-ray cavities is a good indicator of the internal Mach number of AGN jets. Internally subsonic jets produce bottom-wide cavities, while internally supersonic jets tend to produce non-bottom wide cavities, including center-wide, top-wide, or cylindrical cavities. This result is very robust, as projection effect, jet size and duration do not affect top-wideness appreciably.

We looked at two archetypical radio-mode AGN feedback events in Perseus \citep{reynolds15} and Cygnus A \citep{fabian00}. The two X-ray cavities in Cygnus A appear to be center-wide, suggesting that they were produced by light internally-supersonic jets. On the other hand, the two inner cavities in Perseus are bottom-wide, indicating that they were possibly produced by internally subsonic jets. These two examples suggest that internally subsonic AGN jets indeed exist and there may be two different types of AGN jets in galaxy clusters divided by the internal Mach number.

Our simulations show that there are subtle differences in the long-term evolution of different types of AGN jets. Very light jets tend to evolve into old cavities significantly elongated along the direction perpendicular to the jet direction (i.e., ``pancakes" described in \citealt{churazov01}), while light jets evolve into old cavities more spread along the jet direction. The outer northwestern X-ray cavity (ghost cavity) in Perseus appears to be a ``pancake" viewed edge-on, suggesting that it was produced by a very light jet. Another possible example of pancake-shaped old cavities is the outer 8-shaped radio lobes in M87 \citep{owen00}, which do not appear to be cavities in deep X-ray maps \citep{million10}. This is not consistent with the picture of double spherical radio lobes, which would have appeared as two clear cavities in X-ray maps. On the other hand, this mystery may be easily explained if the radio lobes are pancakes viewed close to face-on, which produce much less depressions in X-ray surface brightness.

Recent simulations of jet-mode AGN feedback (e.g., \citealt{yuan15}; \citealt{yang16}) often assume that most or all of the material accreted by the central SMBH is ejected in AGN jets, which tends to trigger light or possibly even heavy jets in simulations (though not explicitly stated in these studies). However, this assumption is not consistent with detailed MHD simulations of black hole hot accretion flows, which instead found that almost all the accreted material is ejected in winds (\citealt{yuan15}; also see \citealt{yuan12a} and \citealt{yuan12b}). AGN jets were predicted in these studies to only carry away a small fraction of the accreted mass, and very light jets may be common.

We further point out that very-light internally-subsonic jets may be present in real galaxy clusters, and play an important role in radio-mode AGN feedback. Compared to light internally-supersonic jets often adopted in previous numerical studies, very-light internally-subsonic jets decelerate faster and rise much slower in the ICM, possibly depositing a larger fraction of jet energy to cool cluster cores and thus alleviating the problem of low coupling efficiencies \citep{vernaleo06} without invoking precessing jets or wide jets proposed recently (e.g., \citealt{yang16}; \citealt{hillel16}). 

In our simulations of long-term jet evolution, a significant level of shear viscosity is adopted to suppress interface instabilities and the formation of torus-like structures. Viscosity has also been invoked in the ICM to explain the coherent structure of cold filaments often observed in cool-core galaxy clusters, which would otherwise have been easily disrupted by turbulence (\citealt{fabian03}). A substantial viscosity level in the ICM is also consistent with the stability of cold fronts observed in some galaxy clusters (e.g., \citealt{zuhone14}).

\section*{Acknowledgments}

This work was supported by Chinese Academy of Sciences through the Hundred Talents Program and National Natural Science Foundation of China (Grant No. 11643001). F.G. acknowledges generous support by the Zwicky Prize Fellowship from Swiss Federal Institute of Technology Zurich during the early stages of this project. F.G. thanks Eugene Churazov for the suggestion of making Fig. 2 and Christine Jones for encouragements during the 2015 Snowcluster meeting. F.G. also thanks the referee for a prompt and insightful report.


\label{lastpage}

\end{document}